\documentclass[twocolumn,showpacs,preprintnumbers,amsmath,amssymb]{revtex4}

\usepackage{graphicx}
\usepackage{amsmath,amssymb}
\usepackage{epsfig}
\usepackage{dcolumn}
\usepackage{bm}
\usepackage{rotating}
\usepackage{times}
\usepackage{color}
\newcommand {\be}{\begin{equation}}
\newcommand {\ee}{\end{equation}}
\newcommand {\bea}{\begin{eqnarray}}
\newcommand {\eea}{\end{eqnarray}}
\newcommand {\bem}{\begin{displaymath}}
\newcommand {\eem}{\end{displaymath}}

\newcommand {\f}{\frac }

\begin{document}

\preprint{ }

\title{Size effects in the long-time quasi-static heat transport}

\author{George Y. Panasyuk}
\email{George.Panasyuk.ctr@wpafb.af.mil}
\author{ Kirk L. Yerkes}
\affiliation{Aerospace Systems Directorate,  Air Force Research Laboratory,
 Wright-Patterson Air Force Base, OH 45433}

\date{\today}

\begin{abstract}
We consider finite size effects on heat transfer between thermal reservoirs mediated by a
quantum system, where the number of modes in each reservoir is finite. 
 Our approach is based on the generalized 
quantum Langevin equation and the thermal reservoirs are described as ensembles of oscillators 
within the Drude-Ullersma model. A general expression for the heat current between the thermal
reservoirs in the long-time quasi-static regime, when an observation time is of the order 
of $\Delta ^{-1}$ and $\Delta$ is the mode spacing constant of a thermal reservoir, is obtained.
The resulting equations that govern the long-time relaxation for the mode temperatures and the
average temperatures of the reservoirs are derived and approximate analytical solutions are found.
The obtained time dependences of the temperatures and the resulting heat current reveal 
peculiarities at $t = 2\pi m/\Delta$ with nonnegative integers $m$ and the heat current vanishes 
non-monotonically when $t \rightarrow \infty$. The validity of Fourier's law for a chain of 
finite-size macroscopic subsystems is considered. As is shown, for characteristic times of 
the order of $\Delta ^{-1}$ the temperatures of subsystems' modes deviate from each other and 
the validity of Fourier's law cannot be established. In a case when deviations of initial 
temperatures of the subsystems from their average value are small, $t \rightarrow \infty$ asymptotic 
values for the mode temperatures do not depend on a mode's number and are the 
same as if Fourier's law were valid for all times.
\end{abstract}

            
\maketitle

\section{Introduction}

One of the most fundamental pursuits in modern physics is the way heat transfers
through  microscopic systems 
(such as nanotubes, molecules, or quantum dots)~\cite{Dhar, Dubi_Di_Ventra}.
Despite of the recent advances, this study still presents many challenges due to
intrinsic non-equilibrium nature of the problem.
Beyond a purely academic interest in the problem, research suggests that nanoscale 
and molecular
systems may be good candidates for many technological advances, such as molecular
wires, molecular diodes, rectifiers, and switches \cite{Jortner, Hanggi}.

In accordance with empirically established Fourier's law, the heat 
flux ${\bf J}$ through both fluids and solids is determined by the expression 
${\bf J} = -\kappa \nabla T({\bf r})$, where the temperature $T$ varies slowly 
on the microscopic scale and $\kappa$ is the thermal conductivity. Despite the 
ubiquitous occurrence of this phenomenon, very few rigorous mathematical derivations of 
this law are known~\cite{Boneto}. While for three-dimensional generic models Fourier's 
law is expected to be true, this law may not be valid for one- and two-dimensional 
systems~\cite{Dhar}. 
The problem acquired nowadays even more attention due
to growing interest in energy transfer at the nanoscale and possible use of nanostructures
for energy applications~\cite{Michel2006,Dubi_DiVentra2009,Chang2008,Chang2006,Chang2007}.

A recently developed approach to study heat transport at the microscopic
level is usually based on the quantum Langevin equation,  
first considered in Ref.~\cite{Senitzky} for a weakly damped harmonic oscillator.
In Ref.~\cite{Mori,Ford}, it was used to formulate transport, collective motion, and
Brownian motion from a unified, statistical-mechanical point of view. 
Later, in Refs.~\cite{Caldeira, Haken, Klimontovich, Allahverdian, Nieuwenhuizen},
the Langevin equation was used for studying the thermalization of a 
particle coupled harmonically to a thermal reservoir and other closely-related problems.
The developed Langevin approach was generalized 
in Refs.~\cite{Zurcher, Saito, Dhar_Shastry, Segal_Nitzan_Hanggi} in order to explore
the steady-state heat current and temperature profiles
in chains of harmonic oscillators placed between two thermal baths, which were
considered as infinitely large, i.e. having infinitely large number of modes. 
An important alternative to the Langevin approach is the non-equilibrium 
Green's function
(NEGF) method. It was developed at first to describe electron transport and calculate the 
steady-state properties of a finite system connected to reservoirs that are modeled by 
non-interacting Hamiltonians with infinite degrees of freedom~\cite{Caroli, Meir, Datta}.
Various important quantities, such as currents and local densities, can be obtained
using the steady-state density matrix and can be written in terms of the Keldysh Green's
functions~\cite{Keldysh}. Later, the NEGF approach was applied to phonon 
transport~\cite{Ozpineci, Yamamoto, Wang1, Wang2, Ojanen, Dhar_Roy2006}. However,  
for non-interacting systems, the Langevin approach reproduces the NEGF results
exactly~\cite{Dhar, Wang_review}.  
Recently, a new method for an exact solution to the Lindblad and Redfield master equations,
which can be also considered as an alternative to the quantum Langevin equation,
has been developed~\cite{Prosen,Prosen1}.

In this paper, we investigate size effects in quasi-static heat transfer between two thermal
 reservoirs described as a {\it finite} collections of quantum harmonic oscillators mediated
by a quantum system, which is also considered in the harmonic approximation.

Study of size effects in nano-structured materials occupy an important part of contemporary 
research. One profound theoretical question is related to the applicability of macroscopic 
theories when a particle has only few nanometers in size.  While study of size and quantum 
effects and their influence on linear and nonlinear response on electromagnetic fields have 
a rather long history (see, for example, 
Refs~\cite{Hache, Rautian, Panasyuk2008, Govyadinov, Panasyuk2011}), systematic 
investigation of the role of these effects and its influence on thermal properties of small 
bodies took part only recently. In Refs.~\cite{ Adiga, Sopu, Pohl}, static thermodynamic 
properties of nanostructures were investigated. In Ref.~\cite{Adiga}, using molecular 
dynamics simulations,  the authors analyzed the local structure and vibrational properties 
of the grain boundary in ultrananocrystal diamond. 
In Ref.~\cite{Sopu}, the authors studied the phonon density of 
states in different nanostructures. They showed that all discontinuities (such as grain 
boundaries and interfaces) introduce vibrational modes with low frequencies that directly 
affect the thermal properties of the material, such as the specific heat. 
In Ref.~\cite{Pohl}, a Monte-Carlo simulations on order-disorder 
transition in 
Pt-Rh nanoparticles were performed in order to study size-dependent trends such as the 
lowering of the critical ordering temperature and the broadening of the compositional 
stability range of the ordered phases. Finally, in Ref.~\cite{Cuansing} the 
authors revealed the critical role of the on-site pinning potential in establishing 
quasi-steady-state conditions of heat transport in finite quantum systems.

Our approach is based on the quantum Langevin equation and employs the Drude-Ullersma
 model for a bath mode spectrum. The developed model allows one to obtain
the heat current between the thermal reservoirs and explore the baths' temperature 
relaxation in the quasi-static regime when an observation 
time can be of the order of the ``Heisenberg'' time $\tau_H \sim \Delta^{-1}$, 
in which case the discreteness of a reservoir's energy spectrum becomes 
resolvable~\cite{Haake, Nieuwenhuizen}. 

The paper is organized as follows. The  model is 
introduced in Sec. II, where the generalized Langevin equation is derived and solved.
This derivation assumes that the number of modes in the thermal reservoirs is finite.
 In Sec. III, expressions for the heat current between the baths and equations that
govern temperature relaxation of the baths' modes are derived. 
An analytical approach for solving the temperature equations, temporal behavior of the 
heat current, and a question of applicability of Fourier's law for a chain of finite 
macroscopic subsystems are considered in Sec. IV. 
Finally, Sec. V provides brief summary to our research.

\section{MODEL}

The total Hamiltonian of the system under consideration is similar to that in 
Refs.~\cite{ Segal_Nitzan_Hanggi, Ford_Lewis_Connell, PLY}:
\be
\label{Htot}
{\mathcal H}_{\rm tot} = {\mathcal H} + {\mathcal H}_{\rm B1} + {\mathcal H}_{\rm B2} + 
 {\mathcal V}_1 + {\mathcal V}_2 .
\ee
Here
\be
\label{H}
\mathcal H = \f{p^2}{2m} +  \f{k x^2}{2}
\ee
is the Hamiltonian of the quantum system (the mediator) described as a harmonic oscillator,
\be
\label{HBnu}
 \mathcal H_{{\rm B}\nu} = \sum_i \left [ \f{p_{\nu i}^2}{2m_{\nu i}} + 
\f{m_{\nu i}\omega_{\nu i}^2x_{\nu i}^2}{2}\right ] 
\ee
are the Hamiltonians of the $\nu$th baths ($\nu$ = 1, 2), and
\be
\label{Vnu}      
 \mathcal V_{\nu} =  -x\sum_i C_{\nu i}x_{\nu i} + 
x^2\sum_i \f{C_{\nu i}^2}{2m_{\nu i} \omega_{\nu i}^2}
\ee
are the Hamiltonians that describe interaction between the mediator and
the baths. 
In Eq. (\ref{H}), $x$ and $p$ are the 
displacement and momentum operators and
$m$ and $k$ are the particle's mass and the spring constant, respectively. 
In Eqs. (\ref{HBnu}) and (\ref{Vnu}), $x_{\nu i}$ and $p_{\nu i}$ are the displacement 
and momentum operators, whereas $m_{\nu i}$ and 
$\omega_{\nu i}$ are the masses and frequencies of the oscillators for the 
$i$th mode that belongs to the $\nu$th bath. Finally, $C_{\nu i}$ are the 
coupling coefficients that describe the interaction between the quantum system and 
the baths. The last contributions to the right hand side of (\ref{Vnu}) are 
self-interaction terms, which guarantee that 
${\mathcal H}_{\rm B\nu} + {\mathcal V}_{\nu}$ is positively defined for 
each $\nu$.

Solutions of the Heisenberg equations 
\be
\label{heis_xnui}
\dot{x}_{\nu i} = \f{p_{\nu i}}{m_{\nu i}}\,\,\,\,{\rm and}\,\,\,\, 
\dot{p}_{\nu i} = -m_{\nu i}\omega_{\nu i}^2x_{\nu i} + C_{\nu i}x 
\ee
for the baths' operators can be presented as
\begin{eqnarray}
\label{heis_xnui_sol}
x_{\nu i}(t) = x_{\nu i}(0)\cos (\omega_{\nu i}t) + 
\f{p_{\nu i}(0)}{m_{\nu i}\omega_{\nu i}}\sin (\omega_{\nu i}t) + \nonumber \\
 \f{C_{\nu i}}{m_{\nu i}\omega_{\nu i}}\int_0^t\sin [\omega_{\nu i}(t-s)]x(s)ds 
\end{eqnarray}
and
\begin{eqnarray}
\label{heis_pnui_sol}
p_{\nu i}(t) = m_{\nu i}\dot{x}_{\nu i}(t) = -m_{\nu i}\omega_{\nu i}x_{\nu i}(0)
\sin (\omega_{\nu i}t) + \nonumber \\
p_{\nu i}(0)\cos (\omega_{\nu i}t) +
 C_{\nu i}\int_0^t\cos [\omega_{\nu i}(t-s)]x(s)ds .
\end{eqnarray}
After substituting (\ref{heis_xnui_sol}) into the other dynamic equations
\be
\label{heis_x}
\dot{x} = \f{p}{m} \,\,\,\,{\rm and}\,\,\,\,\dot{p} = 
-k x + \sum_{i\nu}C_{\nu i}x_{\nu i} - x\sum_{i\nu}\f{C_{\nu i}^2}{m_{\nu i} \omega_{\nu i}^2} ,
\ee
which describe our quantum system, 
one obtains the following quantum Langevin equation: 
\be
\label{langevin}
m\ddot{x} = -k x(t) + \eta (t) - \int_0^t\gamma (t-s)\dot{x}(s)ds - \gamma(t)x(0) ,
\ee
where
\begin{eqnarray}
\label{eta}
\eta (t) = \sum_{i\nu}C_{\nu i}\left [x_{\nu i}(0)\cos (\omega_{\nu i}t) + 
\f{p_{\nu i}(0)}{m_{\nu i}\omega_{\nu i}}\sin (\omega_{\nu i}t)\right ] \,\,\,\,
\end{eqnarray}
is the noise that comes from the baths and
\begin{eqnarray}
\label{gamma}
\gamma (t) = \sum_{i\nu}\f{C_{\nu i}^2}{m_{\nu i}\omega_{\nu i}^2}\cos (\omega_{\nu i}t) 
\end{eqnarray}
is the friction kernel which takes into account the interaction of the quantum 
particle with both thermal reservoirs. 

The Drude-Ullersma model~\cite{Ullersma,  Weiss, Nieuwenhuizen} that we employ here
 assumes that in the absence of the
interaction with the quantum system, each bath consists of uniformly spaced 
modes and introduces the following frequency dependence for the coupling coefficients:
\be
\label{DUM}
\omega_{\nu i} = i\Delta_{\nu}, \,\,\,\, C_{\nu i} = 
\sqrt{\f{2\gamma_{\nu}m_{\nu i}\omega_{\nu i}^2\Delta_{\nu} D_{\nu}^2}{\pi (\omega_{\nu i}^2 +  
D_{\nu}^2)}} 
\ee
where $ i = 1, 2, ... N_{\nu}$. In Eq. (\ref{DUM}), $\Delta_{\nu}$ are the mode 
spacing constants,  $D_{\nu}$  are the characteristic cutoff 
frequencies qualitatively similar to the Debye frequency,  and $\gamma_{\nu}$ are the coupling 
constants 
between a given reservoir  and the mediator. 
Hereafter we assume that the heat baths are identical, which means that
\be
\label{assump} 
X_1 = X_2 \equiv X \,\, {\rm and}\,\, \gamma_1 = \gamma_2 \equiv \gamma/2,
\ee
where $X_{\nu} =N_{\nu}, \,\,\Delta_{\nu}$, or $D_{\nu}$.
However, unlike in Ref.~\cite{PLY}, we consider $\Delta$ as a small but finite parameter. 
In this case, the friction kernel (\ref{gamma}) must be considered as a periodic function with 
a finite period ${\mathcal T} = 2\pi/\Delta$. Using~\cite{Prudnikov}, one finds that
\be
\label{gamma_H}
\gamma (t) = \gamma D[e^{-Dt} + e^{-({\mathcal T}-t)D}] \,\,\,\,{\rm for} \,\,\,\, 
0 \leq t \leq {\mathcal T}
\ee 
and continued periodically with the period ${\mathcal T}$ beyond this interval in accordance 
with the relation
$\gamma (t + {\mathcal T}) = \gamma (t)$. In deriving (\ref{gamma_H}), we approximated the  
finite sum by the 
corresponding series. Due to the fast convergence of (\ref{gamma}), the resulting error is small. 
It also does not change the result qualitatively because the periodicity property is determined 
by the first harmonic in (\ref{gamma}).

Equation (\ref{langevin}) can 
be solved by the Laplace transformation~\cite{Laplace}:
\be
\label{langevin_solved}
x(t) =\dot{g}(t)x(0) + \f{1}{m}g(t)p(0) + \f{1}{m}\int_0^tg(t-s)\eta(s)ds .
\ee
Detailed derivation of (\ref{langevin_solved}) can be found in 
Refs.~\cite{Nieuwenhuizen, Segal_Nitzan_Hanggi}, where similar problems were considered. 
Here $\dot{g} \equiv dg/dt$,
\begin{eqnarray}
\label{g}
g(t) = L^{-1}\left [ \f{1}{z^2 + w_0^2 + z{\hat\gamma}(z)}\right ] = 
\f{1}{2\pi i}\int_{c-i\infty}^{c+i\infty}\f{e^{zt}dz}{h(z)},  \,\,\,\,
\end{eqnarray}
where $L^{-1}$ is the inverse of the  Laplace transform $L$,
\begin{eqnarray}
\label{gamma_hat}
\nonumber
{\hat\gamma}(z) = \f{1}{m} L[\gamma(t)] = 
\f{D\hat\gamma}{1-e^{-z{\mathcal T}}}\times 
\\
\nonumber
\left ( \f{1-e^{-(D+z){\mathcal T}}}{D+z}+
\f{e^{-D{\mathcal T}}-e^{-z{\mathcal T}}}{z-D}\right ) \approx 
\\
\f{D\hat\gamma}{1-e^{-z{\mathcal T}}}\left ( \f{1}{D+z}+\f{e^{-z{\mathcal T}}}{D-z}\right )
\,\,\,\, {\rm with}\,\,\,\hat\gamma = \gamma/m,
\end{eqnarray}
$\omega_0 = \sqrt {k/m}$ is the quantum particle's frequency, and
\be
\label{h}
h(z) = 
z^2 + w_0^2 + \f{D\hat\gamma z}{1-e^{-z{\mathcal T}}}\left ( \f{1}{D+z}+\f{e^{-z{\mathcal T}}}{D-z}\right ) . 
\ee
In Eqs. (\ref{gamma_hat}) and (\ref{h}) we neglected $O[\exp (-D{\mathcal T})]$ terms.

In order to obtain $g(t)$, one can resort to the Heaviside expansion theorem in accordance with which
\begin{eqnarray}
\label{g_final_gen}
g(t) = \sum_ne^{z_nt}\f{1}{h^{\prime}(z_n)}, \,\,\,\,
 h^{\prime}(z) = \f{dh(z)}{dz},
\end{eqnarray}
and $z_n$ are the roots of $h(z)$. The roots can be found iteratively as expansions over 
the small parameter $\Delta$:
\be
\label{root_expan}
z_n = i\omega_n + z_{1n} + z_{2n} + ... \,\equiv i\omega_n + z_n^{\prime}
\ee  
where $\omega_n = n\Delta$, integer $n \geq 1$, and $z_n^{\prime} = z_{1n} + z_{2n} + ...$.
Equation $h(z_n) = 0$ can be written as 
\be
\label{zprime}
(1-e^{-z_n^{\prime}{\mathcal T}})(z_n^2+\omega_0^2)+
\hat\gamma Dz_n\left ( \f{1}{D+z_n}+\f{e^{{-z_n^{\prime}{\mathcal T}}}}{D-z_n}\right )=0 .
\ee
Here we take into account that $\omega_n\mathcal T = 2\pi n$ and 
$\exp(-iz_n\mathcal T)=\exp(-iz_n^{\prime}\mathcal T)$. Solving (\ref{zprime}) 
with respect to $\exp(-iz_n^{\prime}\mathcal T)$, Eq. (\ref{zprime}) can be rewritten as
\be
\label{z_prime}
e^{-z_n^{\prime}{\mathcal T}} = 
\f{\omega_0^2+z_n^2+\hat\gamma Dz_n/(D+z_n)}
{\omega_0^2+z_n^2-\hat\gamma Dz_n/(D-z_n)} .
\ee
Thus, the first correction, $z_{1n}$,
is determined by
\be
\label{z_1n}
e^{-z_{1n}{\mathcal T}} = 
\f{\omega_0^2-\omega_n^2+i\hat\gamma D\omega_n/(D+i\omega_n)}
{\omega_0^2-\omega_n^2-i\hat\gamma D\omega_n/(D-i\omega_n)},
\ee  
where $z_n$ from the right hand side of (\ref{z_prime}) is substituted by its zero order 
approach, $i\omega_n$.
It gives 
\be
\label{z_1n_final}
z_{1n} = -i\Delta\psi (\omega_n ) \equiv -i\Delta\psi_n , 
\ee
where
\begin{eqnarray}
\label{psi}
\psi (\omega ) = 
\f{1}{\pi}{\arctan}
\left [ \f{\hat\gamma D^2\omega}{(\omega_0^2-\omega^2)(D^2+\omega^2)+\hat\gamma D\omega^2}\right ].
\,\,\,\,\,\,
\end{eqnarray}
The second correction, $z_{2n}$, is determined from the same equation (\ref{z_1n}), where
$z_{1n}$ is substituted by $z_{1n} + z_{2n}$ on the left hand side of that equation 
and $\omega_n$ is substituted
by $\nu_n = \omega_n - \Delta \psi_n$ on the right hand side of (\ref{z_1n}). 
As is clear, the resulting equation 
for $z_{2n}$ is $z_{1n} + z_{2n} = -i\Delta\psi (\nu_n )$, or
\begin{eqnarray}
\label{psi_2}
\nonumber
z_{1n} + z_{2n} = -i\Delta [\psi (\omega_n ) - 
\f{\partial \psi (\omega_n)}{\partial \omega_n}\Delta \psi_n ]
+ O(\Delta^3)\\
= z_{1n} + i\Delta^2\f{\partial\psi_n}{\partial \omega_n}\psi_n + O(\Delta^3).
\end{eqnarray}
Thus,
\be
\label{z_2n}
z_{2n} = i\Delta^2\f{\partial \psi_n}{\partial \omega_n}\psi_n = 
z_{1n}O(\tau \Delta),
\ee 
where $\tau$ is a time needed to establish the steady-state heat current. 
In what follows, we assume that $\tau$  satisfies inequality
\be
\label{time_relations}
\tau \equiv {\rm max}(\hat\gamma^{-1}, \omega_0^{-1},  D^{-1}) \ll \Delta^{-1} .
\ee 
In this case, one 
can neglect $z_{2n}$ and Eq. (\ref{g_final_gen}) gives
\begin{eqnarray}
\label{g_final}
\nonumber
g(t) = -\f{i\hat\gamma D^2}{\pi}
\sum_{n=-\infty}^{\infty}\f{\Delta \nu_ne^{i\nu_n t}}{{\rm den}(\nu_n)} = \\
\f{2\hat\gamma D^2}{\pi}\sum_{n \geq 1}\f{\Delta \nu_n\sin (\nu_nt)}{{\rm den}(\nu_n)},
\end{eqnarray}
where $\nu_n = \omega_n - \Delta \psi_n$, ${\rm den}(\nu) = A(\nu)A^*(\nu)$, and
\be
\label{A}
A(\nu) = (D-i\nu)(\omega_0^2-\nu^2)-i\hat\gamma D\nu \, . 
\ee 
As is clear, ${\rm den}(\nu)$ can be rewritten as
\begin{eqnarray}
\label{den_1}
\nonumber
{\rm den}(\nu) = (\omega_0^2-\nu^2)^2D^2+\nu^2(\omega_0^2-\nu^2+\hat\gamma D)^2 = \\
(\nu^2+\mu_1^2)(\nu^2+\mu_2^2)(\nu^2+\mu_3^2),\,\,\,\,\,\,\,\,\,\, 
\end{eqnarray} 
where $\mu_{1,2,3}$ are the roots of equation
\be
\label{3roots}
(D-\mu)(\omega_0^2+\mu^2)-\hat\gamma D\mu = 0 
\ee 
and satisfy an inequality $\Re (\mu_{1,2,3}) > 0$. 

When $\Delta \rightarrow 0$, the sum 
in (\ref{g_final}) transforms to the integral
\begin{eqnarray}
\label{g_old}
g(t) \equiv g_0(t) = -\f{i\hat\gamma D^2}{\pi}
\int_{-\infty}^{\infty}\f{d\nu \nu e^{i \nu t}}
{{\rm den}(\nu)} .
\end{eqnarray}
Taking into account (\ref{den_1}) and closing the integration contour in the upper complex half 
plane (for $t > 0$), one finds the following (expected) result~\cite{Nieuwenhuizen,PLY}:
\be
\label{g_old_final}
g_0(t) = L^{-1}[{\tilde g}(z)] = \sum_{j=1}^3g_je^{-\mu_jt}
\ee
where 
\be
\label{g_tilde}
{\tilde g}(z) = \f{D+z}{(D+z)(z^2+\omega_0^2)+\hat\gamma D z}.
\ee 
Derivation of (\ref{g_old_final}) is facilitated by noting that
\be
\label{g-expansion}
\f{\hat\gamma D^2}{(\nu^2+\mu_1^2)(\nu^2+\mu_2^2)(\nu^2+\mu_3^2)} = \sum_{j=1}^3\f{g_j}{\nu^2+\mu_j^2},
\ee 
which can be also considered as a definition of the coefficients $g_j$'s.
Expression (\ref{g_old_final}) has been obtained and used 
for study fundamental issues of statistical thermodynamics of 
a quantum particle couple to a heat bath in Ref.~\cite{Nieuwenhuizen} when $D$ is large.
We consider here a more general case of the heat transfer between thermal 
reservoirs when 
$D$, $\omega_0$, and $\hat\gamma$ can be comparable, but the relation (\ref{time_relations}) is 
satisfied with finite $\Delta$.

\section{QUASI-STATIC HEAT BALANCE}
\label{heat_flux}

As was shown~\cite{PLY}, the rate of change of the energy of the given $\nu$th thermal reservoir is 
determined by
\be
\label{q_balance}
\f{d}{dt}\sum_{i=1}^N\left \langle  \f{p_{\nu i}^2}{2m_{\nu i}} + 
\f{m_{\nu i}\omega_{\nu i}^2x_{\nu i}^2}{2} \right \rangle =
-\langle {\mathcal P}_{\nu}\rangle ,
\ee 
where the angular brackets denote the ensemble averaging and
\be
\label{flux_nu}
\langle {\mathcal P}_{\nu}\rangle = 
-\sum_{i=1}^N\f{C_{\nu i}}{2m_{\nu i}}\langle p_{\nu i} x + x p_{\nu i}\rangle 
\ee 
is the work that the quantum system performs on the $\nu$th 
bath per unit of time (the  power dispersed in the $\nu$th bath)~\cite{Zurcher}. 
Here $p_{\nu i} = p_{\nu i}(t)$, $x_{\nu i} = x_{\nu i}(t)$, and $x = x(t)$ are 
the solutions (\ref{heis_xnui_sol}), (\ref{heis_pnui_sol}), and  (\ref{langevin_solved}), respectively. 
These solutions, as well as the resulting balance equation (\ref{q_balance}) are accurate in 
the frame of the adopted harmonic approximation. 
Thus, Eq. (\ref{q_balance}) provides a correct description of the energy balance for any 
moment $t \geq 0$. In the general case, (\ref{q_balance}) is a complicated equation because 
it describes both initial transient processes that occur at a microscopic time 
$\tau$ as well as a subsequent long-time quasi-static variation of the 
reservoirs' temperatures. Our goal here is to consider only the long-time quasi-static relaxation 
which happens on a much longer scale $\tau_{\rm H} \sim \Delta^{-1} >> \tau$, 
as was indicated in Ref.~\cite{PLY}. 
In this regard,
we can neglect variations in the baths' temperatures over time intervals of the order of $\tau$.
In this case, after substitution (\ref{heis_pnui_sol}) and  (\ref{langevin_solved}) 
into (\ref{q_balance}), one can also drop all the terms that contain 
explicitly $g(t)$ or $\dot g(t)$. Indeed,  $g(t)$ differs noticeably from zero 
only on the time intervals of the order of $\tau$ near $t = n{\mathcal T}$ where $n \geq 0$ is an 
integer and the corresponding 
contributions cannot influence the temperature variations.
It results in the following expression (see also~\cite{PLY}):
\begin{eqnarray}
\label{Jth_ghi}
\nonumber
\langle {\mathcal P}_{\nu}\rangle  
\approx -\f{1}{2m}\sum_{i=1}^N\f{C_{\nu i}}{m_{\nu i}}[\cos (\omega_{\nu i}t)
\int_0^tdsg(t-s)\times \\
\nonumber
\langle p_{\nu i}(0)\eta (s) + \eta (s) p_{\nu i}(0)\rangle 
 - m_{\nu i}\omega_{\nu i}\sin (\omega_{\nu i}t) \times \\
\int_0^tdsg(t-s)\langle x_{\nu i}(0)\eta (s) + \eta (s) x_{\nu i}(0)\rangle] + J_{\nu}^{\prime},
\,\,\,
\end{eqnarray}   
where
\begin{eqnarray}
\label{J_prime}
\nonumber
J_{\nu}^{\prime} = 
-\f{1}{2m}\sum_{i=1}^N\f{C_{\nu i}^2}{m_{\nu i}}\langle \int_0^td\tau \cos \omega_{\nu i}(t-\tau)x(\tau) 
\times \\
\int_0^tdsg(t-s)\eta (s) +
\nonumber \\ \int_0^tdsg(t-s)\eta (s) \int_0^td\tau \cos \omega_{\nu i}(t-\tau)x(\tau)\rangle .
\,\,\,\,\,\,
\end{eqnarray} 
In the quasi-static (or steady-state) regime,  the power acquired by one reservoir is taken from 
the other, so that $ \langle {\mathcal P}_1\rangle$ = -$\langle {\mathcal P}_2\rangle$.
Thus, we can define the quasi-static heat current in the symmetric form:
\be
\label{q_flux}
J_{\rm th} = \f{1}{2}\langle {\mathcal P}_1 - {\mathcal P}_2\rangle . 
\ee 
If one takes into account (\ref{assump}), 
$J_{\nu}^{\prime}$ in (\ref{Jth_ghi}) is canceled out simplifying the following 
derivation.

In order to find the contributions preceding $J_{\nu}^{\prime}$  in (\ref{Jth_ghi}), 
we use the following approach. As was 
shown in~\cite{Allahverdian,Nieuwenhuizen}, after coupling of a quantum particle to a thermal bath, 
the whole system comes to equilibrium after a microscopic time $\tau$. The thermal bath 
will be comprised of a sum of independent modes having frequencies 
$\nu_k = \omega_k - \Delta \psi (\omega_k)$ where  
$\psi (\omega_k) \equiv \psi_k$ coincides with (\ref{psi}). 
We have now two thermal reservoirs at different temperatures. 
However, one can assume that the influence of the quantum 
particle on both reservoirs is small and each reservoir at any moment of time can be 
characterized by the equilibrium density matrix
\be
\label{rho}
\rho_{\nu} = 
Z_{\nu}^{-1}e^{-\hbar\sum_k \beta_{\nu k}\nu_{\nu k}(n_{\nu k}+1/2)}, 
\ee 
where $ n_{\nu k} = a_{\nu k}^+a_{\nu k}$, 
\be
\label{Z}
Z_{\nu} = {\rm Tr}[e^{-\hbar\sum_k \beta_{\nu k}\nu_{\nu k}(n_{\nu k}+1/2)}] ,
\ee 
and $\beta_{\nu k} = 1/k_{\rm B}T_{\nu k}$,  but we allow now (slow) temperature variations for each 
mode of both reservoirs: $T_{\nu k} = T_{\nu k}(t)$. 
Due to this assumption, one can easily find expressions for 
$\langle x_{\nu i}(0)\eta (s) + \eta (s) x_{\nu i}(0)\rangle$ and 
$\langle p_{\nu i}(0)\eta (s) + \eta (s) p_{\nu i}(0)\rangle$ in (\ref{Jth_ghi}). Using the 
symmetric form (\ref{q_flux}) of the quasi-static heat current, performing the time integrations, 
and employing the Drude-Ullersma model, one finds (see the Appendix)
\be
\label{q_flux_final}
J_{\rm th} = -\f{\hbar \hat\gamma D^2}{2\pi}\sum_{k=1}^N\f{\Delta \nu_k^2}{\nu_k^2+D^2} G(\nu_k,t) 
(n_{1k} - n_{2k}) ,
\ee  
where $n_{\nu k} \equiv n(T_{\nu k},\nu_k) = 1/[\exp (\hbar\nu_k/k_{\rm B}T_{\nu k}) - 1]$ 
are the phonon occupation numbers for each mode of the respective ($\nu$th) thermal reservoir and
\be
\label{G}
G(\nu_k, t) = \int_0^t g(s){\sin}(\nu_ks) ds.
\ee
In the general case, $g(t)$ is determined by Eq. (\ref{g_final}). In a special case, when the 
observation time $t$ is small compared to $\Delta^{-1}$, the mode temperatures $T_{\nu k}(t)$ 
can be considered unchanged during the heat transfer:
\be
\label{Tnuk=Tnu}
T_{\nu k}(t) \approx T_{\nu k}(0) = T_{\nu}.
\ee
Here $T_{1,2}$ are the reservoirs' temperatures before they are interconnected by the 
quantum particle. We assume that each thermal reservoir was 
initially in a state of thermal equilibrium with a particular temperature $T_{\nu}$.
Also, if $\tau \ll t \ll \Delta^{-1}$ (in particular, when $\Delta \rightarrow 0$), 
\be
\label{G_old}
G(\nu_k, t) \approx \int_0^t g_0(s){\sin}(\nu_ks)ds = 
\nu_k\sum_{j=1}^3\f{g_j}{\nu_k^2+\mu_j^2} .
\ee
In this case, one can also replace the sum in (\ref{q_flux_final}) by the corresponding 
integral over the frequency and the resulting heat current
\be
\label{J_old}
J_{\rm th}^{(0)} = -\f{\hbar D^2\hat\gamma}{2\pi}\sum_{j=1}^3 g_j\mu_j^2\int_{0}^{\infty}
\f{d\omega \omega [n(T_1,\omega)-n(T_2,\omega)]}{(D^2+\omega^2)(\mu_j^2+\omega^2)}
\ee
and the corresponding heat conductance $K = J_{\rm th}^{(0)}/\delta T$ at 
$\delta T = T_1-T_2 \rightarrow 0$ reduce to the respective quantities derived in~\cite{PLY}. 

In the case when $\Delta$ satisfies relation (\ref{time_relations}) but is finite and 
$t \sim \Delta^{-1}$, 
one has to return to Eq. (\ref{q_flux_final}). 
In accordance with (\ref{rho}), the average energy of the $\nu$th thermal reservoir 
$\langle E_{\nu}\rangle$ is the sum of the average energies 
$\langle E_{\nu k}\rangle$ of its independent oscillator modes and the time derivative of 
$\langle E_{\nu}\rangle$ is determined as
\be
\label{t_var}
\f{d}{dt}\sum_{k=1}^N\langle E_{\nu k}\rangle = 
\sum_{k=1}^N\f{d}{dt}\f{\hbar\nu_k}{2}{\coth}\left (\f{\hbar\nu_k}{2k_{\rm B}T_{\nu k}}\right ).
\ee 
\begin{figure}                                                                  
\includegraphics[width=6cm]{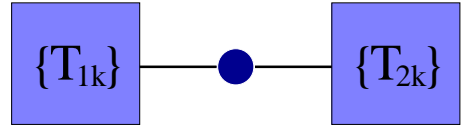}                                      
\caption{\label{bath_2}(Color online) Diagram representation of the energy balance 
(\ref{E-balance_k}).
The squares correspond to the heat baths consisting of independent modes characterized
by  temperatures $T_{1,2k}$. The circle represents the quantum system and the lines stand for the 
interaction between the quantum system and the thermal reservoirs.}
\end{figure}
As follows from (\ref{flux_nu}), (\ref{q_flux}), and (\ref{q_flux_final}), the energy balance 
for the 1st bath is
\begin{eqnarray}
\label{E_first}
\f{d}{dt}\sum_{k=1}^N\langle E_{1k}\rangle = 
-\f{\hbar \hat\gamma D^2}{2\pi}\sum_{k=1}^N\f{\Delta \nu_k^2G(\nu_k,t)}{\nu_k^2 + D^2}  
[n_{1k} - n_{2k}]\,\,
\end{eqnarray}
and is satisfied if each ($k$th) mode satisfies its own energy balance equation
\begin{eqnarray}
\label{E-balance_k}
\nonumber
\f{d}{dt}\langle E_{1k}\rangle
= C(\nu_k, T_{1k})\dot T_{1k} = \,\,\,\,\,\,\,\, \\
-\f{\hbar \hat\gamma \Delta D^2\nu_k^2 }{2\pi(\nu_k^2 + D^2)}G(\nu_k, t) 
[ n(T_{1k},\nu_k)- n(T_{2k},\nu_k)],\,\,\,\,\,\,\,\,
\end{eqnarray} 
which determines the temperature $T_{1k}$ of the $k$th mode. Here $C(\nu_k, T_{1k})$ is the heat 
capacitance of the $k$th mode of the first bath: 
\be
\label{C}
C(\nu_k, T_{1k}) = 
\f{k_{\rm B}}{4}\left (\f{\hbar\nu_k}{k_{\rm B}T_{1k}}\right )^2 
\f{1}{{\sinh}^2(\hbar\nu_k/2k_{\rm B}T_{1k})} .
\ee
The same equation can be written for the 
temperature $T_{2k}$ of the $k$th mode of the of the second thermal reservoir (with the same 
frequency 
$\nu_k$)  by interchanging
$T_{1k}$ and $T_{2k}$ in (\ref{E-balance_k}).
Figure \ref{bath_2} contains graphical illustration of Eq. (\ref{E-balance_k}). 
If $|T_{1k} - T_{2k}| \ll (T_{1k} + T_{2k})/2 \equiv {\bar T}_k$, (\ref{E-balance_k}) 
can be reduced to
\be
\label{dot_Tgen}
C_k(T_{1k})\dot T_{1k} = 
-\f{\hat\gamma\Delta \nu_kD^2}{2\pi(\nu_k^2 + D^2)}G(\nu_k, t)C_k(\bar T_k)(T_{1k} - T_{2k})
\ee 
or, if one neglects the difference between $C_k(T_{1k})\equiv C(\nu_k, T_{1k})$ and 
$C(\nu_k, \bar T_k)$, which is equivalent to dropping $O(T_{1k} - T_{2k})^2$ contributions 
to (\ref{dot_Tgen}), it results in
\be
\label{dot_T1}
\dot T_{1k} = 
-\f{\hat\gamma\Delta \nu_kD^2}{2\pi(\nu_k^2 + D^2)}G(\nu_k, t)(T_{1k} - T_{2k}) .
\ee 
Finally, due to the symmetry between the baths that follows from our assumption (\ref{assump}), 
$T_{1k}(t) + T_{2k}(t) = T_1 + T_2 \equiv 2\bar T$ does not depend on time and (\ref{dot_T1}) 
can be rewritten as
\be
\label{dot_delT}
\f{d}{dt}\delta T_k = -\f{\hat\gamma\Delta \nu_kD^2}{\pi(\nu_k^2 + D^2)}G(\nu_k, t)\delta T_k ,
\ee   
where $\delta T_k \equiv T_{1k} - T_{2k}$.
These equations can  be solved independently for 
each $\nu_k$ with the following initial conditions
\be
\label{bc}
\delta T_k(0) = T_1 - T_2 
\ee 
that are independent on the mode number $k$.

As is clear, the form of Eq. (\ref{dot_delT}) is the same for the classical 
(high-temperature) and quantum (low-temperature) cases. However, for the classical case, 
\be
\label{class}
n(T_{1k},\nu_k) - n(T_{2k},\nu_k) \approx \f{k_{\rm B}}{\hbar \nu_k}(T_{1k} - T_{2k})
\ee  
and Eq. (\ref{dot_delT}) is accurate (no need for any additional 
linearizing to produce
(\ref{dot_Tgen}) and (\ref{dot_delT}) from (\ref{E-balance_k})).

One can also define average temperatures $T_{1,2}(t)$ of each bath from the condition that 
$T_{1,2}(t)$ provide the same total energies of the baths:
\be
\label{aveT_def}
\sum_{k=1}^N\f{\hbar\nu_k}{2}\coth \f{\hbar\nu_k}{2k_{\rm B}T_{\nu k}(t)} =
\sum_{k=1}^N\f{\hbar\nu_k}{2}\coth \f{\hbar\nu_k}{2k_{\rm B}T_{\nu}(t)}, 
\ee  
where $\nu = 1,2$. As is clear,
\be
\label{cond_T12}
T_{\nu}(t=0) = T_{\nu} \,\,\,\,{\rm and}\,\,\,\,T_1(t) + T_2(t) = T_1+T_2 = 2\bar T .
\ee
Subtracting Eq. (\ref{aveT_def}) at $\nu = 2$ from  Eq. (\ref{aveT_def}) at $\nu = 1$,
one finds that
\begin{eqnarray}
\label{diff}
\nonumber
\sum_{k=1}^N\hbar\nu_k[n(\nu_k,T_{1k}(t))-n(\nu_k,T_{2k}(t))] = \\
\sum_{k=1}^N\hbar\nu_k[n(\nu_k,T_1(t))-n(\nu_k,T_2(t))] . 
\end{eqnarray}
If $|T_1(t) - T_2(t)| \ll \bar T$, relation (\ref{diff}) results in
\begin{eqnarray}
\label{Tdiff}
\sum_{k=1}^NC(\nu_k,{\bar T}_k)(T_{1k}-T_{2k}) = 
\delta T(t)\sum_{k=1}^NC(\nu_k,{\bar T}),\,\,\,\,\,\,\,\,
\end{eqnarray}
where $\delta T(t) \equiv T_1(t) - T_2(t)$.  Thus, taking into account that 
${\bar T}_k = {\bar T}$ for each $k$, we find the following expression for
$\delta T(t)$:
\begin{eqnarray}
\label{Tdiff_final}
\delta T(t) = \left [ \sum_{k=1}^NC(\nu_k,\bar T)\right ]^{-1}
\sum_{k=1}^NC(\nu_k,{\bar T})\delta T_k(t) .
\end{eqnarray}
One can expect that these temperatures $T_{1,2}(t)$ can be established in a case when
the  baths have small nonlinearities due to phonon-phonon 
interaction~\cite{Nieuwenhuizen},
provided that a thermalization time $\tau_{\rm therm}$ for the baths' modes 
satisfies the following condition:
\be
\label{tau_therm}
\tau \ll \tau_{\rm therm} \ll \Delta^{-1}.
\ee
In this case, $T_{1,2}(t)$ determine the temperatures of the thermal reservoirs. 
Using (\ref{cond_T12}), one finds
\be
\label{T1,2}
T_{1,2}(t) = \f{1}{2}(T_1+T_2) \pm \f{1}{2}\delta T(t) .
\ee

\subsection{G factor}
\label{subsec:coupling}

After substituting (\ref{g_final}) into (\ref{G}), 
factor $G \equiv G(\nu_k, t)$ (``G factor'') can be expressed as
\begin{eqnarray}
\label{G_gen} 
G = \f{\hat\gamma}{\pi}\sum_{n\geq 1}\f{ D^2\Delta \nu_n}{{\rm den}(\nu_n)}
\left [ \f{\sin (\nu_k-\nu_n)t}{(\nu_k-\nu_n)} - \f{\sin (\nu_k+\nu_n)t}{(\nu_k+\nu_n)}\right ].
\,\,\,\,
\end{eqnarray}
It is impossible to derive closed forms for $g(t)$ or $G(\nu_k, t)$ because of the 
frequency shift $\Delta \psi_n$. One can find, however, an approximate analytical expression for 
$G(\nu_k, t)$ in the following way. For $t \gg \tau$, one can neglect 
the second contribution in the square brackets in (\ref{G_gen}) and write
\be
\label{G_appr1}
G(\nu_k, t) \approx \f{\hat\gamma \nu_kD^2}{\pi  {\rm den}(\nu_k)}
\sum_{n\geq 1}\f{\Delta \sin (\nu_k-\nu_n)t}{(\nu_k-\nu_n)} .
\ee
The last sum can be easily found~\cite{Prudnikov} and the result is
\be
\label{sum}
\sum_{n\geq 1}\f{\sin (\nu_k-\nu_n)t}{(\nu_k-\nu_n)} = \pi f(\Delta t),
\ee
where 
\be
\label{f}
f(x) = \theta (x) + 2\sum_{m\geq 1}\theta(x - 2\pi m)
\ee
and $\theta (x)$ is Heaviside's $\theta$-function. Finally, taking into account that 
$|\Delta \psi_n| \ll \omega_n$ for any $n \geq 1$ and smallness of 
$\Delta/\omega_0$, (\ref{G_appr1}) can be rewritten as
\begin{eqnarray}
\label{G_final_appr}
G(\nu_k, t) \approx \f{\hat\gamma \nu_kD^2}{{\rm den}(\nu_k)}f(\Delta t) \approx
\f{\hat\gamma \omega_kD^2}{{\rm den}(\omega_k)}f(\Delta t) .
\end{eqnarray}
It is interesting to notice that the same result can be obtained if one simply neglects the 
frequency shift $\Delta \psi_n$ in (\ref{g_final})  and 
(\ref{G}), setting $\nu_n \approx \omega_n$ there. In this case, using 
Eq. (\ref{g-expansion}), taking into account~\cite{Prudnikov}
\begin{eqnarray}
\label{subsid}
\sum_{n=-\infty}^{\infty}\f{\Delta\omega_n\sin (\omega_nt)}{\omega_n \pm i\mu_j} = 
\f{\pi{\rm sinh}[(\pi-\Delta t)a_j]}{{\rm sinh}(\pi a_j)},
\end{eqnarray}
where $a_j = \mu_j/\Delta$, and neglecting small quantities $|e^{-\mu_j{\mathcal T}}|$,
 one finds that
\begin{eqnarray}
\label{g(t)final}
g(t) = \sum_{j=1}^3g_j[e^{-\mu_jt} - e^{-({\mathcal T}-t)\mu_j}],\,\,\,\,{\rm if}\,\,\,\,
0 \leq t \leq {\mathcal T}
\,\,\,\,\,\,
\end{eqnarray}
and is continued periodically for $t > {\mathcal T}$ in accordance with the relation
\be
\label{period}
g(t+{\mathcal T}) = g(t).
\ee
Using 
(\ref{g(t)final}), (\ref{period}), and the definition (\ref{G}) with $\nu_n \approx \omega_n$,
one arrives at the same result (\ref{G_final_appr}). 
Strictly speaking, as one can also understand from this alternative derivation, 
(\ref{G_final_appr}) is correct only if 
$m{\mathcal T} + \tau < t < (m+1){\mathcal T} - \tau$, so, again,
our approximation can be applied only if (\ref{time_relations}) is satisfied.

\begin{figure}                                                                  
\includegraphics[width=7.0cm]{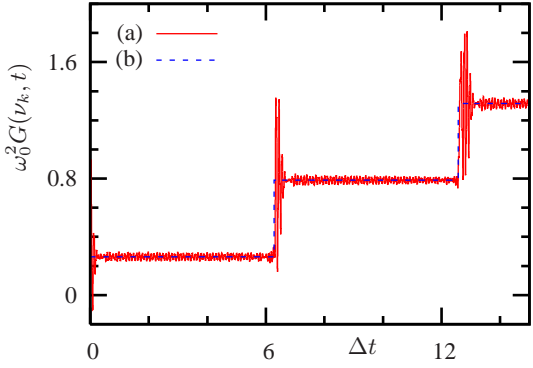}                             
\caption{\label{G_Delta0.0107k50}(Color online) Dependence of the dimensionless $G$ factor on time
when $\hat\gamma/\omega_0 = 0.5$, $D/\omega_0 = 1$, $\Delta/\omega_0 = 0.01$, 
and $\nu_k/\omega_0 = 0.5$.
(a) accurate $G(\nu_k,t)$ determined by (\ref{G_gen}) and (b) its approximation 
(\ref{G_final_appr}).}
\end{figure}
\begin{figure}                                                                  
\includegraphics[width=7.0cm]{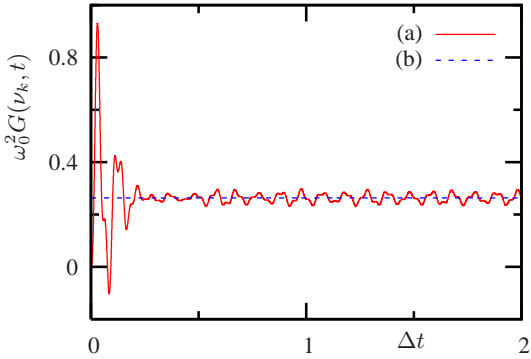}                      
\caption{\label{G_Delta0.0107k50detail}(Color online) Same dependencies as in 
Fig. \ref{G_Delta0.0107k50} 
when $0 \leq \Delta t \leq 2$.}
\end{figure}
\begin{figure}                                                                  
\includegraphics[width=7.0cm]{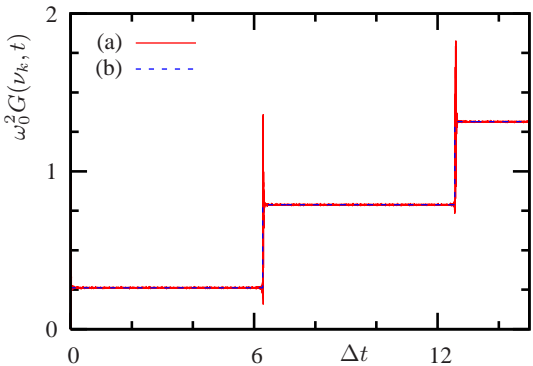}                      
\caption{\label{G_Delta0.00107k500}(Color online) Same dependencies as in 
Fig. \ref{G_Delta0.0107k50}
for $\hat\gamma/\omega_0 = 0.5$, $D/\omega_0 = 1$, $\Delta/\omega_0 = 0.001$, 
and $\nu_k/\omega_0 = 0.5$.}
\end{figure}
\begin{figure}                                                                  
\includegraphics[width=7.0cm]{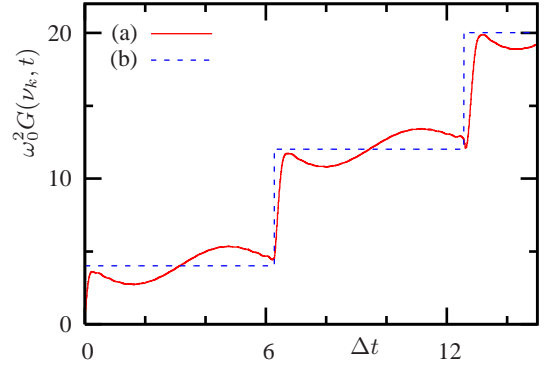}             
\caption{\label{G_Delta0.0107k93}(Color online) Same dependencies as in Fig. \ref{G_Delta0.0107k50}
for $\hat\gamma/\omega_0 = 0.5$, $D/\omega_0 = 1$, $\Delta/\omega_0 = 0.01$, 
and $\nu_k = \nu_{\rm res} \approx \omega_0$.}
\end{figure}

\begin{figure}                                                                  
\includegraphics[width=7.5cm]{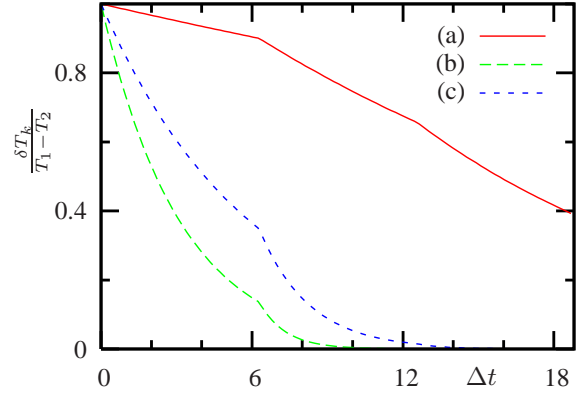}                               
\caption{\label{longTD0.001k}(Color online) Temperature relaxation curves produced by solving
Eq. (\ref{dot_delT_w}) at 
$\hat\gamma/\omega_0 = 0.5$, $D/\omega_0 = 1$, and $\Delta/\omega_0 = 0.001$. 
(a) $\nu_k = 0.5\omega_0$, (b) $\nu_k = \nu_{\rm res} \approx \omega_0$, and 
(c) $\nu_k = 1.1\omega_0$.}
\end{figure}

\begin{figure}                                                                  
\includegraphics[width=7.0cm]{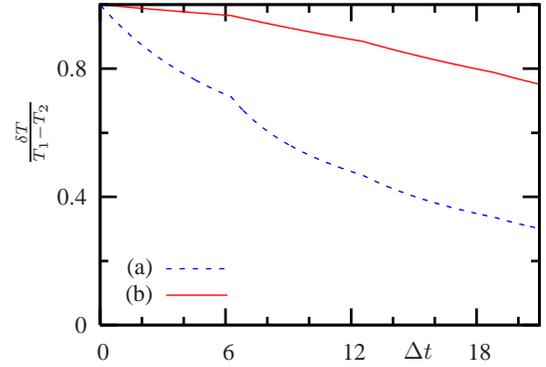}                        
\caption{\label{delT_Delta0.001Dhat1}(Color online) 
Average temperature relaxation curves when $\hat\gamma/\omega_0 = 0.5$, 
$D/\omega_0 = 1$, and $\Delta/\omega_0 = 0.001$. 
(a) $\hbar\omega_0/2k_{\rm B}{\bar T} = 0.1$ and (b) $\hbar\omega_0/2k_{\rm B}{\bar T} = 5$.}
\end{figure}

Figures \ref{G_Delta0.0107k50} - \ref{G_Delta0.0107k93} illustrate results of comparison 
between the 
accurate dimensionless $G$ factor $\omega_0^2G(\nu_k,t)$ determined by (\ref{G_gen}) and the 
corresponding 
approximate analytical expression (\ref{G_final_appr}). 
In all the figures, $\hat\gamma/\omega_0 = 0.5$ and $D/\omega_0 = 1$. The latter choice 
can be explained in the following way. Parameter $D$ is loosely associated to the Debye frequency, 
providing a smooth algebraic frequency cutoff, when the bath
 frequency spectrum
 does not end exactly at $\omega = D$. The number $N$ of the baths'
modes is finite now and we assumed that 
\be
\label{N}
N = \f{\omega_{\rm max}}{\Delta} \sim \f{D}{\Delta}\,\, ,
\ee 
where $\omega_{\rm max}$ is the maximum frequency in the bath spectrum.
When $D \ll \omega_0$, the mediating particle is effectively
uncoupled from the baths~\cite{PLY}. It cannot be excited, and, as a consequence, cannot absorb 
or transfer energy between the thermal reservoirs. 
For this reason, one can assume that $D \sim \omega_0$
 and $\omega_{\rm max} \gtrsim D$. 
On the other hand, in order to
avoid using too large $N$ in the case of small $\Delta$ (which is needed to satisfy
 (\ref{time_relations})), it is desirable also to have $D \lesssim \omega_0$. 
Thus, any values for $D$ and $\omega_{\rm max}$ that satisfy inequality 
$D \lesssim \omega_0 \lesssim \omega_{\rm max}$ are acceptable and we choose here
$D = \omega_0$.
Taking the above observations into account, we choose also $\omega_{\rm max} = 1.3D$. 
In this case, in Figs. \ref{G_Delta0.0107k50} - \ref{G_Delta0.0107k50detail} and in 
Fig. \ref{G_Delta0.0107k93},  we use $\Delta/\omega_0 = 0.01$ and $N = 130$, and in 
Fig. \ref{G_Delta0.00107k500} and Figs. \ref{longTD0.001k} - \ref{delT_Delta0.001Dhat1},
$\Delta/\omega_0 = 0.001$ and $N=1300$ are chosen. As our numerical analysis shows, the
result of summation in (\ref{G_gen}) does not depend noticeably on $N$ provided that
$N > 1.2D/\Delta$ due to the fast convergence.

In Figs. \ref{G_Delta0.0107k50} - 
\ref{G_Delta0.00107k500},  $\nu_k = 0.5\omega_0$ and in Fig. \ref{G_Delta0.0107k93}  
$\nu_k = \nu_{\rm res} \approx \omega_0$. The frequency $\nu_{\rm res}$  is chosen from
a condition that $\nu_{\rm res}$ minimizes den($\nu$):
\be
\label{res}
\f{d}{d\nu}{\rm den}(\nu )|_{\nu=\nu_{\rm res}} = 0.
\ee
 The latter means that the energy exchange between the thermal reservoirs is carried out
by the modes that are in ``resonance'' with the mediating quantum system. 

As is clear from 
Figs. \ref{G_Delta0.0107k50} - \ref{G_Delta0.00107k500}, formula (\ref{G_final_appr}) represents 
fairly well the exact result (\ref{G_gen}) for 
$G(\nu_k , t)$ and the accuracy improves proportionally to $\Delta$.
Indeed, when $\Delta/\omega_0$ 
decreases from 0.01 to 0.001, the relative difference between the accurate and approximate results 
also decreases by approximately ten times. 
The same results are observed for all other ratios $\nu_k/\nu_{\rm res} \neq 1$. 
It is important to notice that deviations of $G(\nu_k , t)$ from its 
approximate analytical expression are not only small for $\Delta/\omega_0 \lesssim 0.01$ 
but also appear as oscillations that occur on a time scale 
of the order of $\tau$. 
In accordance with our assumption, temperature variations are insensitive to these short-scale 
oscillations, and we have an additional argument by which the latter can be neglected. 
In these cases, analytical formula
(\ref{G_final_appr}) can be used for solving equations (\ref{dot_delT}) or 
(\ref{E-balance_k}) for temperature variations. 

A reason why our approximation (\ref{G_final_appr}) works well can be explained 
in the following way. As is clear from Eq. (\ref{psi}), 
$|\psi_k| \leq 1/2$ for all values of its parameters and $|\psi_k| \approx 1/2$ only
when $\nu_k \approx \nu_{\rm res} \approx \omega_0$ and is small otherwise.
When $\nu_k \approx \nu_{\rm res} \approx \omega_0$,
Eq. (\ref{G_final_appr}) may not be accurate,  which is illustrated in Fig. \ref{G_Delta0.0107k93}. 
The main reason for this is because
the factor $\nu/{\rm den}(\nu)$ in (\ref{G_gen}) changes also sharply with $\nu$ when
$\nu \approx \nu_{\rm res}$ and
our approximation (\ref{G_appr1}) can be only qualitatively correct. 
In this case one cannot use (\ref{G_final_appr}) for accurate calculations 
and must employ  (\ref{G_gen}) for solving the temperature equations.
As follows from our numerical analysis, however, the number of $\nu_k$ in the vicinity 
of $\nu_{\rm res}$, which makes Eq. (\ref{G_final_appr}) inaccurate, is relatively small
and decreases with $\Delta$. For example, 
if $\Delta/\omega_0 = 0.001$ and $\hat\gamma/\omega_0$ = 0.5, 
$k_{\rm res} = [\nu_{\rm res}/\Delta]$ = 1130 and formula (\ref{G_final_appr}) already 
provides a reasonable accuracy if $|k - k_{\rm res}| \geq 4$.
 
The step-wise time dependence of $G(\nu_k , t)$, approximated by  
(\ref{G_final_appr}), 
is due to the finiteness of $\Delta$. 
Indeed, in the opposite
case, when $\Delta \rightarrow 0$, we have the steady-state result (\ref{G_old}), 
corresponding to the contribution of the only first term in (\ref{f}).

As follows from our numerical analysis, 
the major contribution to $G(\nu_k , t)$ is determined by the first term in the square brackets
in (\ref{G_gen}). It also comes from a region of the frequency spectrum of $g(t)$ close to
$\nu_k$ but not only from one frequency $\nu_n = \nu_k$. 
The contribution from $\nu_n = \nu_k$ is linear in $t$. The other part
of the sum in (\ref{G_appr1}) is the periodic function of period ${\mathcal T}$ 
and is equal to $\pi - \Delta t$
on each time interval $m{\mathcal T} < t < (m+1){\mathcal T}$ ($m$ = 0, 1, 2, ...).
Thus, the sum (\ref{G_appr1}) results in the step-wise time dependence
determined by (\ref{f}).

For smaller $\hat\gamma$, $\tau \sim \hat\gamma^{-1}$. In this case, our model will be valid for
the proportionally smaller values of $\Delta$ in order to satisfy the inequality 
(\ref{time_relations}) and will produce results similar
 to the ones shown above.

\begin{figure}                                                                  
\includegraphics[width=8.0cm]{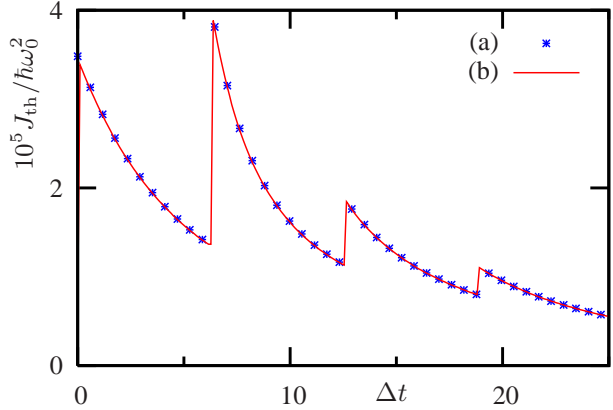}     
\caption{\label{long_gen_Jth_tauhat1}(Color online) 
Dimensionless thermal current curves when $\hat\gamma/\omega_0 = 0.5$, 
$D/\omega_0 = 1$, $\Delta/\omega_0 = 0.001$,  $\hbar\omega_0/2k_{\rm B}{\bar T} = 1$, and 
(a) approximation (\ref{G_final_appr}) is used; (b) accurate expression (\ref{G_gen}) is used.}
\end{figure}

\section{Temperature relaxation and Fourier's law}
\label{Temp_Fourier}

As follows from the above analysis, Eq. (\ref{G_final_appr}) is accurate
for all modes (except $\nu_k = \nu_{\rm res}$) provided that $\Delta$ is sufficiently 
small ($\Delta/\omega_0 \lesssim 0.001$ in the considered example). 
In this case, Eq. (\ref{dot_delT}) can be rewritten as
\be
\label{dot_delT_w}
\f{d}{dt}\delta T_k = -(2 m + 1)\hat \Omega_k\Delta \delta T_k 
\ee
on each time interval $m {\mathcal T}\leq t \leq (m+1){\mathcal T}$, where $m$ = 0, 1, 2, ... and
\be
\label{Omega}
\hat \Omega_k = \f{(\hat\gamma\omega_kD^2)^2}{\pi (\omega_k^2+D^2){\rm den}(\omega_k)}  .
\ee 
Assuming continuity in the temperature variations and taking into account initial 
conditions (\ref{bc}),
solution of Eq. (\ref{dot_delT_w}) can be presented as
\be
\label{delT_sol}
\delta T_k(t) = (T_1 - T_2)e^{-2\pi m^2\hat \Omega_k}e^{-(2m+1)\hat \Omega_k(\Delta t-2\pi m)},
\ee
where $m = m(t) = [t\Delta/(2\pi )]$ is the integer part of the value inside the square brackets.
Using this result, temperature dependencies for the $k$th mode of the $\nu$th thermal reservoir
can be presented as
\be
\label{T_k1,2}
T_{1,2k}(t) = \f{1}{2}(T_1+T_2) \pm \f{1}{2}\delta T_k(t) .
\ee
Figure \ref{longTD0.001k} shows results of application of (\ref{delT_sol}) for several modes. 
As one finds,
the fastest relaxation (leveling of the corresponding temperatures of both reservoirs) occurs
when $\omega_k \approx \omega_0$ due to the resonance character of the heat transport at these 
frequencies (curve (b)). 
The rate of heat exchange decreases as $|\omega_k - \omega_0|$ increases, as also follows 
from the figure (curves (a) and (c)). Due to different relaxation rates, $T_{1k}(t)$ 
(or $T_{2k}(t)$) will all differ when $t \sim \Delta^{-1}$, in accordance with (\ref{delT_sol})
and (\ref{T_k1,2}). Thus, neither of the two thermal reservoirs can be characterized by a 
single local (in time) temperature ($T_1(t)$ or $T_2(t)$) if $t \sim \Delta^{-1}$.

Figure \ref{delT_Delta0.001Dhat1} shows time variations of $\delta T/(T_1-T_2)$ that
represent $\delta T_k/(T_1-T_2)$ averaged over the baths' modes
in accordance with Eq. (\ref{Tdiff_final}) for different values of ratio 
$R = \hbar \omega_0/2k_{\rm B}\bar T$. When $R \ll 1$ (curve (a)), 
we have the high-temperature limit (classical case). As our calculations reveal, when
$R$ decreases below 0.1, all such curves approach to 
$\delta T_{\rm cl}(t) = N^{-1}\sum_{k=1}^N\delta T_k(t)$, as is expected. 
At low temperatures
(quantum regime), when $R$ is large (curve (b)), 
the temperature relaxation is slow. 
This is in accordance with the fact
that the thermal conductance decreases when $\bar T$ decreases.
Prominent features that appear in 
Figs. \ref{longTD0.001k} and \ref{delT_Delta0.001Dhat1} are the 
peculiarities that occur at $t = m{\mathcal T}$ with $m$ = 1, 2, ... . This is a 
consequence of the finite values of $\Delta$ (or ${\mathcal T} = 2\pi/\Delta$, 
see the comment at the end the previous section).

Finally, Fig. \ref{long_gen_Jth_tauhat1} shows the time dependence of the dimensionless heat
current $10^5J_{\rm th}(t)/\hbar \omega_0^2$, where $J_{\rm th}(t)$ is determined by 
(\ref{q_flux_final}). The shown time dependences are generic for the considered model. 
In computing the heat current, we used expression (\ref{T_k1,2}) 
for the temperatures of the baths' modes and assumed that 
$\alpha \equiv (T_1 - T_2)/\bar T$ is small. 
Here $\bar T = (T_1+T_2)/2$ and $T_{1,2}$ are the initial temperatures of the baths.
As our numerical analysis indicates, if $|\alpha | \lesssim 0.01$, linearizing of Eq.
(\ref{E-balance_k}) that gives (\ref{dot_delT}) is well justified and $J_{\rm th} \sim \alpha $.
The factor $G(\nu _k, t)$, which grows stepwise, is suppressed by the exponentially decaying difference of
the phonon occupation numbers for each $\nu_k$ due to the above result (\ref{delT_sol}), 
and $J_{\rm th}(t) \rightarrow 0$ when $t \rightarrow \infty $. 
As we also found (see Fig. \ref{long_gen_Jth_tauhat1}), when $\Delta/\omega_0 \lesssim 0.001$ 
approximate expression (\ref{G_final_appr}) for the G factor gives essentially the same result
for $J_{\rm th}(t)$ as when the corresponding accurate expression (\ref{G_gen}) is used.
Thus, the short-scale oscillations from accurate $G(\nu_k, t)$ around its approximate value 
(\ref{G_final_appr}) average out due to summation in (\ref{q_flux_final}) and smooth resulting
$J_{\rm th}(t)$ is determined by (\ref{G_final_appr}).
It must be mentioned that even when $\Delta t$ is large, the heat current still can 
be non-zero if one recovers the contributions to $J_{\rm th}(t)$ containing 
explicitly $g(t)$ or $\dot g(t)$ (see the text just before Eq. (\ref{Jth_ghi})).
These contributions, however, are the short-scale oscillations that occur during microscopic times 
of the order 
of $\tau$ near $t = n{\mathcal T}$, where $n \geq 0$ is an integer. Because $\tau \ll \Delta ^{-1}$,
where $\Delta ^{-1}$
is the characteristic time scale in Fig. \ref{long_gen_Jth_tauhat1}, we did not consider them in this study.

\subsection{Fourier's law}
\label{subsec:Fourier}

\begin{figure}                                                                  
\includegraphics[width=8.5cm]{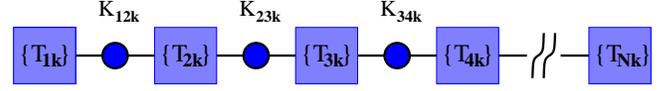}                                    
\caption{\label{bath_N}(Color online) Diagram representation for a chain of
nanoparticles (squares) interconnected by the mediating quantum systems (circles).}
\end{figure}

We consider now a chain of $P$ macroscopic subsystems coupled
by the mediators described by the Hamiltonian (\ref{H}), which is illustrated in Fig. \ref{bath_N}.
Each subsystem and coupling are described by the 
Hamiltonian (\ref{HBnu}) and interaction (\ref{Vnu}), respectively, within the framework of
the Drude-Ullersma model (\ref{DUM}). The extended in this way model reduces to our initial model
(\ref{H}) - (\ref{Vnu}) when $P$ = 2. 
Assume that before the connection, all subsystems are prepared in the
state of thermal equilibrium having temperatures $T_{0n} \equiv T_n(0)$, where 
$n$ = 1, 2, ..., $P$ and
$|T_{0n} - T_{0(n-1)}| \ll \bar T_{0n} \equiv [T_{0n} + T_{0(n-1)}]/2$.  
After a short time $\tau_P$ of initial transient processes 
(we assume that $\tau_P \ll \Delta^{-1}$), 
one can consider this chain as an example of a system in local
thermal equilibrium, consisting of weakly interacting parts (interaction is only due to 
the mediating quantum systems) having temperatures $T_n$ close to $T_{0n}$. 
One can expect that the temperatures at $t > \tau_{\rm P}$
will change in accordance 
with Fourier's law and in the stationary state, achieved when $t \rightarrow \infty$, a uniform 
temperature distribution will be reached~\cite{Boneto}. However, we show here that Fourier's law
cannot be validated in the considered model for $t \sim \Delta^{-1}$.

Indeed, the energy conservation law
can be written in the form similar to Eq. (\ref{dot_Tgen}) for each mode. In particular, 
for the first and $P$th subsystems they read
\begin{eqnarray}
\label{dot_Tgen_1}
C(\omega_k, T_{1k})\dot T_{1k} = - K(\omega_k, {\bar T}_{2k})(T_{1k} - T_{2k})
\end{eqnarray}
and
\begin{eqnarray}
\label{dot_Tgen_P}
C(\omega_k, T_{Pk})\dot T_{Pk} = - K(\omega_k, {\bar T}_{Pk})[T_{Pk} - T_{(P-1)k}],
\end{eqnarray}
respectively, and for the $n$th subsystem, where $2 \leq n \leq P-1$, we have
\begin{eqnarray}
\label{dot_Tgen_n}
\nonumber
C(\omega_k, T_{nk})\dot T_{nk} = K(\omega_k, {\bar T}_{nk})[T_{(n-1)k} - T_{nk}] \\
- K(\omega_k, {\bar T}_{(n+1)k})[T_{nk} - T_{(n+1)k}].
\end{eqnarray}
Here
\be
\label{K_nk}
K(\omega_k,\bar T_{nk}) \approx (m+1/2)\hat\Omega_k\Delta C_k(\omega_k, \bar T_{nk})
\ee
and $\bar T_{nk} = [T_{(n-1)k} + T_{nk}]/2$. These equations can be rewritten in the differential form 
if one introduces a continuous coordinate $x$ = $nd$, where $d$ is the distance between two adjacent 
subsystems. In this case, (\ref{dot_Tgen_n}) can be rewritten as
\begin{eqnarray}
\label{h_balance1}
C_k(T_k)\dot T_k(x)= K_k(x-d/2)[T_k(x-d) - T_k(x)] - \nonumber \\
 K_k(x+d/2)[T_k(x) - T_k(x+d)],\,\,\,\,\,\,\,
\end{eqnarray}
which leads to the energy conservation law for each $k$ mode:
\begin{eqnarray}
\label{Fourier_k}
\tilde C_k(T_k)\dot T_k(x)= \partial_x[\kappa_k(x)\partial_xT_k(x)],
\end{eqnarray}
where $\tilde C_k = C_k/d$ and $\kappa_k = K_kd$ are the specific heat and thermal conductivity
of the $k$th mode of the chain, respectively.

If one considers evolution of the system on a time scale $t \ll \Delta^{-1}$, one can 
approximate the mode temperatures $T_{nk}$ as $T_{nk} \approx T_{0n} \approx T_n$ and
${\bar T}_{nk} \approx {\bar T}_{0n} \approx {\bar T}_n$ with accuracy $O(\Delta t)$. Here we
define a subsystem's temperature $T_n = T_n(t)$ as $T_{nk}$ averaged over the $k$ modes 
in accordance to Eq. (\ref{Tdiff_final}):
\begin{eqnarray}
\label{Tnk_ave}
T_n(t) = \left [ \sum_{k=1}^NC(\nu_k,T_{n0})\right ]^{-1}
\sum_{k=1}^NC(\nu_k,T_{n0}) T_{nk}(t) .
\end{eqnarray}
With the same accuracy, one can also approximate $C(\omega_k, T_{nk}) \approx C(\omega_k, T_n)$ and 
$K(\omega_k,\bar T_{nk}) \approx K(\omega_k,\bar T_n)$.  
Equations (\ref{dot_Tgen_1}) - (\ref{dot_Tgen_n}) can be rewritten now as
\begin{eqnarray}
\label{dot_Tn_1}
C(\omega_k, T_1)\dot T_1 = - K(\omega_k, {\bar T}_2)(T_1 - T_2),
\end{eqnarray}
\begin{eqnarray}
\label{dot_Tn_P}
C(\omega_k, T_P)\dot T_P = - K(\omega_k, {\bar T}_P)[T_P - T_{P-1}],
\end{eqnarray}
and
\begin{eqnarray}
\label{dot_Tn_n}
\nonumber
C(\omega_k, T_n)\dot T_n = K(\omega_k, {\bar T}_n)[T_{n-1} - T_n] \\
- K(\omega_k, {\bar T}_{n+1})[T_n - T_{n+1}],
\end{eqnarray}
respectively. Summing up each of Eqs. (\ref{dot_Tn_1}) - (\ref{dot_Tn_n})
over all $k$ modes of the system, one arrives at
\begin{eqnarray}
\label{dot_T_1_sum}
C(T_1)\dot T_1 = - K({\bar T}_2)(T_1 - T_2),
\end{eqnarray}
\begin{eqnarray}
\label{dot_T_P_sum}
C(T_P)\dot T_P = - K({\bar T}_P)(T_P - T_{P-1}),
\end{eqnarray}
and
\begin{eqnarray}
\label{dot_Tgen_n_sum}
\nonumber
C_{n}(T_n)\dot T_n = \,\,\,\,\,\,\,\,\,\,\,\,\,\, \\
K({\bar T}_n)(T_{n-1} - T_n) - K({\bar T}_{n+1})(T_n - T_{n+1}),\,\,\,\,
\end{eqnarray}  
respectively. Here
\be
\label{K_n_and_C_n}
K({\bar T}_n) = \sum_{k=1}^NK(\omega_k, {\bar T}_n),\,\,\,
C(T_n) = \sum_{k=1}^NC(\omega_k, T_n).
\ee
Thus, the above system (\ref{dot_T_1_sum}) - (\ref{dot_Tgen_n_sum}) is correct if 
one neglects the contributions of the order of $O(\Delta t)$. 
In this case, it can be recast in the 
form of Fourier's law in the same way as is described above:
\begin{eqnarray}
\label{Fourier}
\tilde C(T)\dot T(x)= \partial_x[\kappa(x)\partial_xT(x)],
\end{eqnarray}
which coincides with Eq. (83) from Ref.~\cite{PLY}. For longer times, when $t \sim \Delta^{-1}$, 
temperatures of different modes may deviate significantly from each other, as is
discussed in the text following Eq. (\ref{T_k1,2}). 
It means that thermal equilibrium in each subsystem shown in 
Fig. \ref{bath_N} breaks with time 
and Fourier's law cannot be validated on the time scale $t \sim \Delta^{-1}$. 
As we already discussed, a possible way to restore Fourier's law
is to introduce a weak phonon-phonon interaction which may thermalize our subsystems if
the condition (\ref{tau_therm}) is satisfied. 

It is interesting to notice that in a case when deviations of $T_n$ from their average at
$t = 0$ are small, asymptotic values $T_{nk}(t)$ of our system (\ref{dot_Tgen_1}) - (\ref{dot_Tgen_n})
at $t \rightarrow \infty$ coincide with the corresponding asymptotic values $T_n(t)$ of 
(\ref{dot_T_1_sum}) - (\ref{dot_Tgen_n_sum}) as if Fourier's law were correct at all 
times $t > \tau_{\rm P}$. Indeed, assume that 
\be
\label{small_dev_T_n}
{\rm max}[T_{0n}] - {\rm min}[T_{0n}] \ll {\bar T} \equiv \f{1}{P}\sum_{n=1}^PT_{0n}.
\ee 
In this case, one can approximate
\begin{eqnarray}
\label{C_K}
\nonumber
C(\omega_k, T_{nk}) \approx C(\omega_k, {\bar T}) \equiv C_k, \\
K(\omega_k, {\bar T}_{nk}) \approx K(\omega_k, {\bar T}) \equiv K_k,
\end{eqnarray}
and the system (\ref{dot_Tgen_1}) - (\ref{dot_Tgen_n}) can be rewritten as 
\be
\label{small_1}
C_k\dot T_{1k} = -K_k(T_{1k}-T_{2k}),
\ee
\be
\label{small_P}
C_k\dot T_{Pk} = -K_k[T_{Pk}-T_{(P-1)k}],
\ee
and
\be
\label{small_n}
C_k\dot T_{nk} = -K_k[2T_{nk}-T_{(n+1)k}-T_{(n-1)k}].
\ee
As one can easily find after summing up Eqs. (\ref{small_1}) - (\ref{small_n}),
\be
\label{sum_T}
\sum_{n=1}^P\dot T_{nk} = 0, \,\,\,\, {\rm or} \,\,\,\,\sum_{n=1}^PT_{nk}(t) = C,
\ee
where $C$ is an arbitrary constant. At the same time, as follows already from 
(\ref{dot_Tgen_1}) - (\ref{dot_Tgen_n}), $T_{nk}(t)$ reach the following asymptotic
at $t \rightarrow \infty$:
\be
\label{T_nk_asymp}
T_{1k} = T_{2k} = ... = T_{Pk} \equiv T_k.  
\ee 
Thus, using (\ref{sum_T}) at $t = 0$ and $t = \infty$, one can write
\be
\label{T_kfinal}
\sum_{n=1}^PT_{nk}(0) = C = PT_k, \,\,\,\,{\rm or}\,\,\,\,T_k = {\bar T}.
\ee 
On the other hand, when (\ref{small_dev_T_n}) is satisfied and if Fourier's law were correct, 
one can rewrite system (\ref{dot_T_1_sum}) - (\ref{dot_Tgen_n_sum}) in the same way:
\begin{eqnarray}
\label{dot_T_1_sum_1}
C\dot T_1 = - K(T_1 - T_2), \,\,\,\,C\dot T_P = - K(T_P - T_{P-1}),
\end{eqnarray}
and
\begin{eqnarray}
\label{dot_Tgen_n_sum_1}
C\dot T_n = -K(2T_n - T_{n+1} - T_{n-1}),
\end{eqnarray}  
where
\be
\label{K_n_and_C_n_1}
C \approx C({\bar T}) \,\,\,\, {\rm and} \,\,\,\, 
K \approx K({\bar T}).
\ee
Summing up equations (\ref{dot_T_1_sum_1}) - (\ref{dot_Tgen_n_sum_1}), one finds as before that
\be
\label{asym_1}
\sum_{n=1}^PT_n(0) = PT, \,\,\,\,{\rm or}\,\,\,\, T = \f{1}{P}\sum_{n=1}^PT_n(0)
= {\bar T},
\ee
where $T_1 = T_2 = ... = T_P \equiv T$ is the asymptotic solution of 
(\ref{dot_T_1_sum}) - (\ref{dot_Tgen_n_sum}) when $t \rightarrow \infty$. Thus, the asymptotic 
for mode temperatures
\be
\label{T_k=barT}
T_k = T = {\bar T}
\ee
does not depend on $k$ and coincides with the asymptotic of 
(\ref{dot_T_1_sum}) - (\ref{dot_Tgen_n_sum}), which is a discrete version of 
Fourier's law (\ref{Fourier}).

The chain system discussed above is similar to some extend to chain systems considered
 in Refs.~\cite{Michel2005, Dubi_Diventra}. In Ref. \cite{Michel2005}, a weakly coupled chain
of many-level identical subsystems is explored. 
Each subsystem has a band of $n \sim 1000$ exited states equally distributed
 over a bandwidth $\delta \epsilon \ll \Delta E$, where 
$\Delta E$ is the energy gap between the band and a  non-degenerate ground state.
As was shown, if some conditions on the system Hamiltonian are met
(in particular, if $\lambda \ll \delta \epsilon \ll \Delta E$, where $\lambda$ is a coupling 
constant that describes interaction between neighboring subsystems), 
Fourier's law can be validated. As was also shown by numerical integrating the 
Schr\"{o}dinger equation, no diffusive transport results
if these conditions are violated.
In Ref. \cite{Dubi_Diventra}, a chain also consisting of identical subsystems $S_n$ 
($n$ = 1, 2,  ... , $P$) is attached by its left-most ($S_1$) and right-most ($S_N$) 
subsystems  to 
external environments (very large thermal reservoirs) held at fixed temperatures $T_L$ and $T_R$,
 respectively. It is assumed that there are no interactions in subsystems $S_n$. 
As is shown, however, due to the contacts $S_1$ and $S_N$ with the corresponding
 environments, Fourier's law can be established in the case of weak enough interaction 
between $S_n$'s. 
This result is also in accordance with Ref. \cite{Wu}, where 
Fourier's law is derived for a general system that satisfies the same conditions as the 
model~\cite{Dubi_Diventra} does. 
    A model that always exhibits validity of Fourier's law is the model of self-consistent 
reservoirs~\cite{Vissher1975, Dhar_Roy2006, Malay_Segal2011}. It is constructed from a harmonic 
chain of quantum particles placed between two large thermal reservoirs (like in the 
model~\cite{Dubi_Diventra}) by connecting each quantum 
particle 
(subsystem) to a heat bath. Temperatures of these baths are determined by demanding that there 
is no net heat current between the chain and these reservoirs in the steady-state. The introduction 
of self-consistent thermal reservoirs introduces dephasing in the system's dynamics and leads 
inherently to local equilibrium and onset of Fourier's law. This is in contrast to the 
models~\cite{Michel2005,Dubi_Diventra}, where the validity of Fourier's law is not guaranteed 
depending on the chain Hamiltonian. In particular, it breaks down if the coupling between 
the subsystems is strong~\cite{Dubi_Diventra}. 

All these examples, including our chain system, show that the fact that a system is
one-dimensional alone does not mean that Fourier's law is violated. The validity (or violation)
of Fourier's law depends on the Hamiltonian which underlines the type of system's interactions 
and their strength.

\section{Conclusions}
\label{conclusion}

We have considered the finite-size effects in heat transport between two
heat baths mediated by a quantum particle using the generalized quantum Langevin equation. 
Both heat baths and the quantum system are considered 
in the harmonic approximation. We derive expressions for the quasi-static heat 
current for the case when each thermal reservoir comprises of a finite number of modes
having a finite mode spacing $\Delta$. In the limiting case when $\Delta \rightarrow 0$,
the previously obtained expressions for the steady-state heat current 
and the corresponding heat conductance are restored.
The resulting equations that govern long-time ($t \gtrsim \Delta^{-1}$) relaxation
for the mode temperatures and the average temperatures of the baths are derived 
and solved. 
Time dependencies of these temperatures as well as the heat current show 
peculiarities
at $t = 2\pi m/\Delta$, where $m$ = 1, 2, ... due to finite  $\Delta$.
In particular, the heat current decays to zero in a non-monotonic fashion.
The solutions depend on a small number of measurable parameters, such as 
the frequency of the quantum particle, the coupling constant, and the Debye 
cutoff frequency. 
It is important to notice that recently a new techniques employing quantum dots
as temperature probes for measuring the temperature of a nanoparticle has been 
developed~\cite{Gupta}. The temperature information is inferred from the fluorescent 
intensity of the quantum dots. 
As the temperature increases, the maximum intensity of the 
fluorescent signal shifts toward larger wavelengths and its magnitude decreases.
Either of these two changes may be used to find nanoparticle's temperature.
This techniques has the potential to verify predictions of our model.

The validity of Fourier's law 
for a chain of the finite-size identical subsystems is discussed.
On a short time scale, when $t \ll \Delta^{-1}$,  we return to the case
considered in Ref.~\cite{PLY} where Fourier's law was validated. 
When $t \sim \Delta^{-1}$, the temperatures of different baths' modes deviate
from each other preventing thermal equilibrium in each subsystem and
the validity of Fourier's law cannot be established. 
As is found, when deviations of the initial subsystems' temperatures $T_n$ 
from their average value $\bar T$ are small,
the $t \rightarrow \infty$ asymptotic values of the mode temperatures  do not depend 
on the mode number and have the same value $\bar T$ as in the case
if Fourier's law were valid for all times.

\section*{ACKNOWLEDGMENTS}
The authors wish to acknowledge that this research was funded by the Air Force 
Office of Scientific Research and the National Research Council Senior Associateship 
Award at the Air Force Research Laboratory. We acknowledge valuable discussions with
Dr. G. A. Levin.

\section*{APPENDIX}
\appendix
\setcounter{section}{1}

Taking into account our assumption (\ref{assump}), one can drop the index $\nu$ from 
the frequencies of the 
baths' modes, and the dynamical variables $x_{\nu i}(t)$,  $p_{\nu i}(t)$, 
and $\eta(t)$ are determined by the following expressions~\cite{Nieuwenhuizen}:
\begin{eqnarray}
\label{2st_way_x}
x_{\nu i}(t) = \sum_{k=0}\sqrt{\f{\hbar}{2m_{\nu i}\nu_k}}e_i^k
(a_{\nu k}^+e^{i\nu_{k}t} + a_{\nu k}e^{-i\nu_{k}t}), \,\,\,\,\,\,\,\,\,\,\,\,\,\,
\end{eqnarray}
$p_{\nu i}(t) = m_{\nu i}\dot{x}_{\nu i}(t)$, and
\begin{eqnarray}
\label{eta_noise}
\nonumber
\eta(t) = \sum_k \sqrt{\f{\hbar \gamma \nu_k \Delta D^2}{2\pi (D^2+\nu_k^2)}}  \\
\times [e^{i(\phi_k+\nu_k t)}a_k^+ + e^{-i(\phi_k+\nu_k t)}a_k], \,\,\,\phi_k \equiv \pi \psi_k .
\end{eqnarray}
Here $e_i^k$ are orthonormal 
eigenvectors corresponding to the $k$th mode~\cite{Nieuwenhuizen}, which are determined by 
\begin{eqnarray}
\label{e_ik}
e_i^k =
\f{2\Delta \omega_i \sin [\phi (\omega_k)]}{\pi (\omega_i^2 - \nu_k^2)}
 \sqrt{\f{ D^2+\nu_k^2}{D^2+\omega_i^2}}.
\end{eqnarray}
Using these expressions in the averages 
$\langle x_{\nu i}(0)\eta (s) + \eta (s) x_{\nu i}(0)\rangle$ and
$\langle p_{\nu i}(0)\eta (s) + \eta (s) p_{\nu i}(0)\rangle$, the formulas for the Bose 
occupation numbers
\be
\label{bose}
\langle a_k^+a_{k_1} +  a_{k_1}a_{k}^+\rangle = 
{\coth}(\hbar \nu_k\beta_k/2) \delta_{k k_1} 
\ee
and for the averages $\langle a_k a_{k_1}\rangle$ = $\langle a_k^+ a_{k_1}^+\rangle$ = 0,  
one can find
\begin{eqnarray}
\label{ave_x}
\nonumber
\langle x_{\nu i}(0)\eta (s) + \eta (s) x_{\nu i}(0)\rangle =
\sum_k\sqrt{\f{\hbar^2\gamma \Delta D^2}{4\pi m_{\nu i}(D^2+\omega_i^2)}}
\,\,\,\,\, \\
 \times\f{4\omega_i \sin (\phi_k) \Delta}{\pi (\omega_i^2 - \nu_k^2)}
\coth(\beta_k\hbar \nu_k/2)\cos(\nu_k t + \phi_k)\,\,\,\,\,\,\,\,\,\,\,\,
\end{eqnarray}
and
\begin{eqnarray}
\label{ave_p}
\nonumber
\langle p_{\nu i}(0)\eta (s) + \eta (s) p_{\nu i}(0)\rangle =
\sum_k\sqrt{\f{\hbar^2\gamma \nu_k^2  m_{\nu i}\Delta D^2}{4\pi (D^2+\omega_i^2)}} 
 \,\,\,\,\,\,\, \\
\times\f{4\omega_i \sin (\phi_k) \Delta}{\pi (\omega_i^2 - \nu_k^2)}
\coth(\beta_k\hbar \nu_k/2)\sin(\nu_k t + \phi_k) .\,\,\,\,\,\,\,\,\,\,\,
\end{eqnarray}
Substituting (\ref{ave_x} and (\ref{ave_p}) in Eq. (\ref{Jth_ghi}) and using that
\begin{eqnarray}
\label{t_int_1}
\nonumber
\int_0^t g(t-s) \cos(\nu_k s + \phi_k)ds = 
\\ \,\,
G_1(\nu_k,t)\cos(\nu_k t + \phi_k)+G(\nu_k,t)\sin(\nu_k t + \phi_k)\,\,\,\,\,\,\,\,\,\,\,
\end{eqnarray}
and
\begin{eqnarray}
\label{t_int_2}
\nonumber
\int_0^t g(t-s) \sin(\nu_k s + \phi_k)ds = 
\\ \,\,
G_1(\nu_k,t)\sin(\nu_k t + \phi_k)-G(\nu_k,t)\cos(\nu_k t + \phi_k),\,\,\,\,\,\,\,\,\,\,\,\,\,
\end{eqnarray}
where
\begin{eqnarray}
\label{G1}
G_1(\nu_k,t) = \int_0^t g(s)\cos(\nu_k s) ds
\end{eqnarray}
and 
\begin{eqnarray}
\label{G2}
G(\nu_k,t) = \int_0^t g(s)\sin(\nu_k s) ds ,
\end{eqnarray}
 one finds
\begin{eqnarray}
\label{ave_P}
\nonumber
\langle {\mathcal P}_{\nu}\rangle = \f{\hbar \hat\gamma\Delta D^2}{\pi^2}
\sum_k \sin (\phi_k)\coth (\beta_{\nu k}\hbar \nu_k/2)\times \,\,\,\,\,\,\,\,\,\,
\\ 
\nonumber
 \{\nu_k[G_1\sin(\nu_k t + \phi_k)-G\cos(\nu_k t + \phi_k)] S_{1k} -\,\,\,\,\,\,\,\,\,\,\,\,
\\ \,\,
[G_1\cos(\nu_k t + \phi_k)+G\sin(\nu_k t + \phi_k)] S_{2k}\}+J_{\nu}^{\prime}. \,\,\,\,
\,\,\,\,\,\,\,\,\,\,\,\,
\end{eqnarray}
In (\ref{ave_P}), $S_{2k}=-\dot S_{1k}$ and $S_{1k} = -\partial_t^2S_k$ with
\begin{eqnarray}
\label{S}
S_k =\sum_i\f{\Delta\cos(\omega_it)}{(\omega_i^2-\nu_k^2)(D^2+\omega_i^2)}= 
\f{A_k - B}{D^2+\nu_k^2}.
\end{eqnarray} 
Using~\cite{Prudnikov}, sums $A_k$ and $B$ are determined as
\be
\label{Ak}
A_k(t) = \sum_i\f{\Delta\cos(\omega_it)}{\omega_i^2-\nu_k^2}=
\f{\Delta}{2\nu_k^2} + \f{\pi \cos(\nu_k t+\phi_k)}{2\nu_k\sin(\phi_k)}
\ee
and
\be
\label{Bk}
B(t) = \sum_i\f{\Delta\cos(\omega_it)}{D^2+\omega_i^2}= 
-\f{\Delta}{2D^2} + \f{\pi\gamma(t)}{2D^2\gamma},
\ee
where $\gamma(t)$ is defined by (\ref{gamma_H}). Thus, $S_k$, $S_{1,2k}$ can
be found. Taking into account (\ref{assump}), (\ref{q_flux}), and relation 
$\coth (\hbar\nu_k\beta_{\nu k}/2) = 1 + 2n_{\nu k}$, $J_{\nu}^{\prime}$ cancels out and (\ref{ave_P})
results in
\be
\label{P_ave_accu}
J_{\rm th} = -\f{\hbar \hat\gamma D^2}{2\pi}
\sum_k \f{\Delta \nu_k^2G(\nu_k,t)}{D^2+\nu_k^2}(n_{1k}-n_{2k})+\delta J_{\rm th}.
\,\,\,\,\,\,\,\,
\ee
Here
\be
\label{delta_J}
\delta J_{\rm th} = \f{\hbar \hat\gamma D^3}{2\pi}[\gamma_+(t)\Gamma_+(t)-\gamma_-(t)\Gamma_-(t)]
\ee
with
\begin{eqnarray}
\label{Gamma+}
\nonumber
\Gamma_+(t) = \sum_{k=1}^N\f{\Delta\nu_k\sin(\phi_k) (n_{1k}-n_{2k})}{\nu_k^2+D^2} \times \\
(G_1(\nu_k,t)\sin (\nu_k t+\phi_k) - G(\nu_k,t)\cos (\nu_k t+\phi_k)), \,\,\,\,\,\,\,\,\,\,
\,\,\,\,
\end{eqnarray}
\begin{eqnarray}
\label{Gamma-}
\nonumber
\Gamma_-(t) = D\sum_{k=1}^N\f{\Delta\sin(\phi_k) (n_{1k}-n_{2k})}{\nu_k^2+D^2} \times \\
(G_1(\nu_k,t)\cos (\nu_k t+\phi_k) + G(\nu_k,t)\sin (\nu_k t+\phi_k)), \,\,\,\,\,\,\,\,\,\,
\,\,\,\,
\end{eqnarray}
and $\gamma_{\pm}(t) = \exp(-Dt) \pm  \exp[-({\mathcal T}-t)D]$ for $0 \le t \le \mathcal T$
and continued periodically with the period $\mathcal T$ beyond this interval. 
As is clear, $\gamma_{\pm}(t)$ are non-zero essentially only within the intervals 
$n\mathcal T - \tau \lesssim t \lesssim n\mathcal T + \tau$. Also, due to the factor 
$n_{1k}-n_{2k}$, the short-scale oscillations from $\delta J_{\rm th}$ are decaying with time and
we neglect $\delta J_{\rm th}$ which results in (\ref{q_flux_final}).

\end{document}